\documentclass[prb,letterpaper,twocolumn,showpacs]{revtex4}

\usepackage{amsbsy,amssymb,amsmath,bm}
\usepackage{graphicx,color,epsfig,rotate}

\begin{document}

\topmargin -15mm
\preprint{APS/123-QED}

\title{Heat capacity of the site-diluted spin dimer system Ba$_3$(Mn$_{1-x}$V$_x$)$_2$O$_8$}

\author{E. C. Samulon, M. C. Shapiro, I. R. Fisher}

\affiliation{Geballe Laboratory for Advanced Materials and Department of Applied Physics, Stanford University, Stanford, California 94305, USA, and Stanford Institute for Materials and Energy Sciences, SLAC National Accelerator Laboratory, 2575 Sand Hill Road, Menlo Park, California 94025, USA.}

\begin{abstract}

Heat capacity and susceptibility measurements have been performed on the diluted spin dimer compound Ba$_3$(Mn$_{1-x}$V$_x$)$_2$O$_8$.  The parent compound Ba$_3$Mn$_2$O$_8$ is a spin dimer system based on pairs of antiferromagnetically coupled $S=1$, 3$d^2$ Mn$^{5+}$ ions such that the zero field groundstate is a product of singlets.  Substitution of non-magnetic $S=0$, 3$d^0$ V$^{5+}$ ions leads to an interacting network of unpaired Mn moments, the low temperature properties of which are explored in the limit of small concentrations, $0\leq x  \leq0.05$.  The zero-field heat capacity of this diluted system reveals a progressive removal of magnetic entropy over an extended range of temperatures, with no evidence for a phase transition.  The concentration dependence does not conform to expectations for a spin glass state.  Rather, the data suggest a low temperature random singlet phase, reflecting the hierarchy of exchange energies found in this system.

\end{abstract}


\maketitle

\section{Introduction}

Randomness can lead to intriguing magnetic states not typically available to perfectly ordered systems. The archetypal example is the spin glass state, found for a wide variety of disordered materials with either site or bond randomness \cite{Mydosh_1993}. In contrast, for gapped systems with a singlet ground state, substitution of non-magnetic elements can introduce local moments, ultimately leading to long range magnetic order due to the effective interaction mediated by the background singlet state - one manifestation of ``order by disorder'' (OBD). For example, both the Spin-Peierls compound CuGeO$_3$ and the spin ladder compound SrCu$_2$O$_3$ have gapped ground states without long range order; however diluting either system with a small amount of non-magnetic Zn or Mg ions onto the $S = \frac{1}{2}$ Cu site induces antiferromagnetic order \cite{Oseroff_1995, Masuda_1998, Ohsugi_1999}. It is not clear whether this effect should be uniformly anticipated for all gapped systems, motivating both theoretical and experimental interest aimed at exploring the properties of disordered spin-gap materials.

The broad category of spin dimer compounds provides a simple means to study the effect of non-magnetic substitution on a singlet ground state. Comprising pairs of spins with a dominant antiferromagnetic nearest neighbor exchange, the ground state is a product of singlets.  The gap to excited triplet states can be closed by an applied field, leading to canted XY antiferromagnetic order \cite{Giamarchi_2008}. Recently, quantum Monte Carlo (QMC) simulations have been performed for the specific case of spin dimers arranged on a square lattice with antiferromagnetic nearest neighbor and next nearest neighbor exchange \cite{Roscilde_2005, Roscilde_2006}.  These calculations revealed that for an appropriate range of concentrations, substitution of non-magnetic impurities leads to long range order in zero magnetic field.  The predicted wave-vector is the same as that found for the stoichiometric parent compound subjected to fields above the critical field. It would be highly desirable to experimentally test whether such an OBD state is found for a real material conforming to the simple effective spin Hamiltonian used in this calculation. Unfortunately, there are not currently any suitable candidate materials that match these requirements for which substitution of non-magnetic species is possible over an appreciable range of concentrations. Conversely, it would be equally interesting to see how geometric frustration affects the stability of the OBD phase. Here, unfortunately, quantum Monte Carlo simulations are prohibitively difficult due to the frustration induced sign problem, and we must resort to experiment to provide insight. In this case, though, we are much more fortunate in that there are several candidate materials to which we can turn. One such material is Ba$_3$(Mn$_{1-x}$V$_x$)$_2$O$_8$.  In this paper, we present the results of an initial survey of the low temperature properties of this material via heat capacity measurements. We find no evidence for a sharp phase transition into an ordered state down to our base temperature of 50 mK.  Rather, the magnetic entropy is found to be removed over an extended range of temperatures, nominally independent of impurity concentration for the range of compositions studied.  These results do not conform to expectations for a canonical spin glass, leading to the possibility that Ba$_3$(Mn$_{1-x}$V$_x$)$_2$O$_8$ manifests instead a random singlet phase at low temperature.

\begin{figure}
\includegraphics[width=8.5cm]{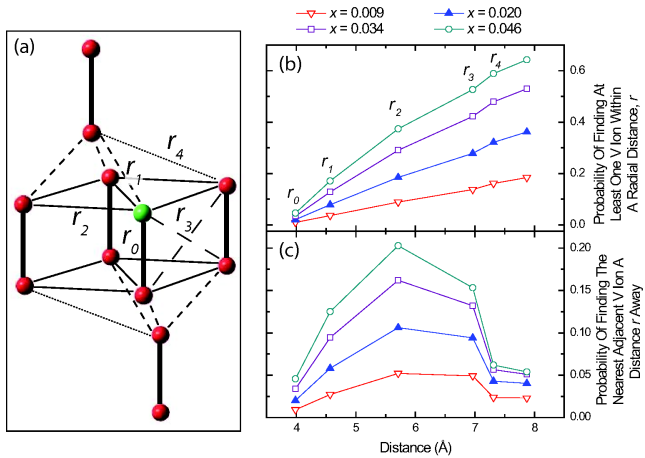}
\caption{(Color online) (a) Schematic diagram showing the transition metal sublattice of the dimer compound Ba$_3$(Mn$_{1-x}$V$_x$)$_2$O$_8$. Mn ions are shown in red. Dominant antiferromagnetic exchange $J_0$ between pairs of Mn moments in the undiluted compound ($x$=0) leads to a singlet ground state. In this diagram a single Mn ion has been substituted by a non-magnetic V ion (green). The Mn ion vertically below the V ion is now no longer paired and hence contributes a local magnetic moment. Distances between the V ion and its nearest neighbors $r_0$ - $r_4$ are labeled. (b) Probability, assuming no clustering, of finding at least one V ion within a distance $r$ from the substituted ion, shown for the specific vanadium concentrations ($x$) studied, and for the specific distances $r_0 - r_4$ \cite{Prob1}  (c) Probability of finding the nearest neighboring V ion a distance $r$ from the substituted ion \cite{Prob2}. }
\label{Struct}
\end{figure}

Ba$_3$Mn$_2$O$_8$ is a spin dimer compound composed of vertical pairs of Mn$^{5+}$ ions arranged on triangular layers (Fig. \ref{Struct}a)) \cite{Weller_1999, Uchida_2002, Tsujii_2005}.  Antiferromagnetic exchange between the $S=1$, 3$d^2$ Mn$^{5+}$ ions forming the dimer leads to a zero field ground state which is a product of singlets with excited triplet and quintuplet states.  Interactions between dimers broadens the excited states, and application of a magnetic field can close the gap to the excited states, leading to three distinct ordered states at high fields \cite{Samulon_2008, Samulon_2009, Suh_2009, Samulon_2010}.  The interactions of this system have been determined through a combination of inelastic neutron scattering (INS), electron paramagnetic resonance (EPR) and thermodynamic measurements of the phase diagram.  INS studies of Ba$_3$Mn$_2$O$_8$ revealed a spin gap of $\Delta=1.05$ meV and a dominant exchange within a dimer (along the $r_0$ direction) as $J_0=1.61$ meV \cite{Stone_2008a, Stone_2008b}.  These studies also determined the nearest and next nearest out-of-plane interdimer exchanges, $J_1$ = 0.118(2) meV and $J_4$ = 0.037(2) meV; and the dominant in-plane interdimer exchanges $J_2 - J_3$ = 0.1136(7) meV ($r_1$ - $r_4$ in Fig. \ref{Struct})(a).  Measurements of the critical fields then yield estimates for $J_2$ and $J_3$ of 0.256 meV and 0.142 meV, respectively \cite{Stone_2008b}.  Finally, EPR measurements of the Mn$^{5+}$ moments in Ba$_3$(Mn$_{1-x}$V$_x$)$_2$O$_8$ for nominal composition of $x=0.75$ revealed a nearly isotropic $g$-tensor, with $g_{aa}$ = 1.96 and $g_{cc}$ = 1.97, and an easy axis single ion anisotropy $D$ = -0.024 meV \cite{Whitmore_1993}.  Similar measurements in the pure Ba$_3$Mn$_2$O$_8$ compound revealed a zero field splitting of the triplet states characterized by $D$ =- 0.032 meV \cite{Hill_2007}.

Ba$_3$V$_2$O$_8$ is isostructural to Ba$_3$Mn$_2$O$_8$.  However in contrast to the strongly magnetic Mn$^{5+}$ ion, V$^{5+}$ corresponds to a 3$d^0$ electron configuration and hence is non-magnetic.  Partial substitution of V in Ba$_3$Mn$_2$O$_8$ therefore leads to unpaired Mn moments (Fig. \ref{Struct}(a)) \cite{Manna_2009}.  For the relevant V concentrations studied here, the data do not suggest clustering of impurities, in which case, for the highest dilution studied, $x=0.046$, the probability is 0.64 to find at least one neighboring V ion within 8 \textrm{\AA} of any given V impurity (Fig. \ref{Struct}(b)).  Furthermore, the most likely distance between nearest neighboring V ions is the $r_2$ pairing (Fig. \ref{Struct}(c)).

\section{Experimental Methods}

Single crystals of Ba$_3$(Mn$_{1-x}$V$_x$)$_2$O$_8$ ($0\leq x \leq 1$) were grown from a NaOH flux, following the procedure that we have previously described for Ba$_3$Mn$_2$O$_8$ \cite{Samulon_2008}.  Polycrystalline precursor was made by mixing BaO, Mn$_2$O$_3$ and V$_2$O$_5$ in appropriate ratios.  Electron microprobe measurements using Ba$_3$Mn$_2$O$_8$ and Ba$_3$V$_2$O$_8$ standards were used to determine the vanadium content of the single crystals.  Measured V concentrations were close to nominal values (Fig. \ref{Suscept}(c)).  Data are presented here for samples with V concentrations 0.009(1), 0.020(1), 0.034(1), 0.046(1) and 0.980(1).  Uncertainties reflect the standard deviation between multiple measurements performed at different locations for individual crystals.  Systematic uncertainties for these low concentrations are likely slightly larger.  Crystals of Ba$_3$Mn$_2$O$_8$ have an intense dark green color.  Crystals of Ba$_3$V$_2$O$_8$ are perfectly clear due to the absence of 3$d$ electrons.  All samples were found to be insulating regardless of vanadium concentration.

Low-field susceptibility measurements were performed using a commercial Quantum Design MPMS XL SQUID magnetometer for fields of 1000 Oe applied perpendicular to the $c$-axis. Heat capacity ($C_p$) studies were performed with a Quantum Design physical properties measurement system (PPMS) using standard thermal relaxation-time calorimetry. These measurements were performed in temperatures down to 50 mK and fields up to 0.5 T parallel to the $c$ axis.

\section{Results}

\begin{figure}
\includegraphics[width=8.5cm]{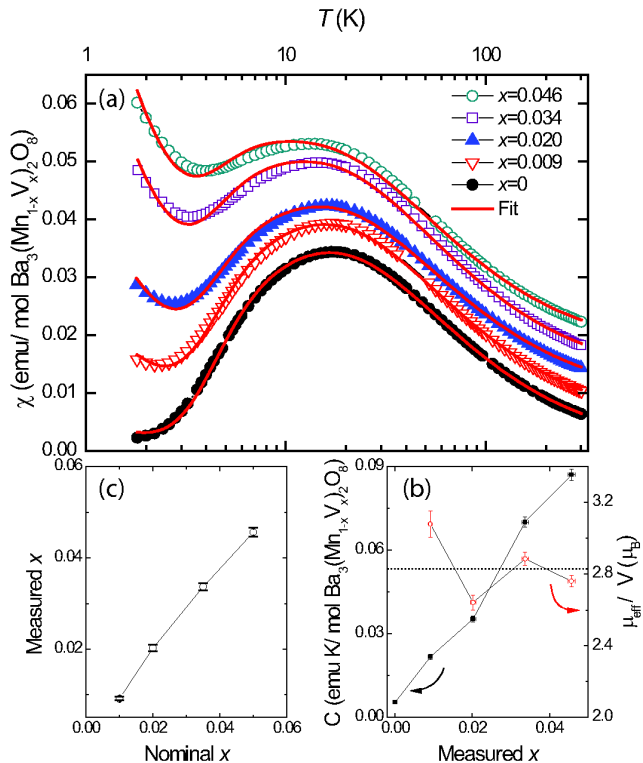}
\caption{(Color online) (a) Susceptibility as a function of temperature for single crystals of Ba$_3$(Mn$_{1-x}$V$_x$)$_2$O$_8$ (compositions listed in legend). Successive curves are offset by 0.004 emu/mol for clarity.  Data are fit (red lines) by a dimer model, described in the main text, including a mean-field correction to account for interdimer interactions, a temperature-independent term, and a Curie-Weiss term to account for the unpaired magnetic impurities introduced by V-substitution. Values for the exchange constants were held fixed at the values obtained for the stoichiometric compound (black circles). (b) Curie constant, C (left axis), and the associated effective moment per V, $\mu_{eff}$/ V (right axis). Note that V$^{5+}$ is non-magnetic, and that the moment arises from the unpaired Mn$^{5+}$ $S$=1 spin on the broken dimers. Dotted horizontal line shows the anticipated effective moment $\mu_{eff} = g\sqrt{S(S+1)}\mu_{B} = 2.83$ $\mu_{B}$ for Mn$^{5+}$ (c) Measured versus nominal composition of the single crystals used in this study.}
\label{Suscept}
\end{figure}

The low field susceptibility of Ba$_3$(Mn$_{1-x}$V$_x$)$_2$O$_8$ for $0\leq x\leq 0.046$ is shown in Fig. \ref{Suscept}(a). Results were fit to a dimer model including a mean field correction, a term corresponding to Curie-Weiss paramagnetic behavior and a temperature independent term, according to the previously described model \cite{Uchida_2002, Samulon_2008}:

\begin{eqnarray}
\chi_{iso} & = & \frac{2N_A\beta g^2 \mu_B^2 \left(1+5e^{-2\beta J}\right)}{3+e^{\beta J} + 5e^{-2 \beta J}}
\nonumber\\
\chi_{total} & = & \alpha \frac{\chi_{iso}}{1+\lambda\chi_{iso}} + \frac{\textrm{C}}{T-\theta} + \chi_0
\label{SusEq}
\end{eqnarray}

Here $\lambda = 3 \left[J_1+J_4+2\left(J_2+J_3\right) \right] / N_A g^2 \mu_B^2$ is a mean field correction to account for interdimer exchange.  $N_A$ is Avogadro's number, $\beta=1/k_B T$ and $\alpha$ is the number of dimers per mole (where 1 mole refers to the formula unit Ba$_3$(Mn$_{1-x}$V$_x$)$_2$O$_8$, such that $\alpha=1$ for $x=0$).  Recent INS studies found a negligible concentration dependence of both the spin gap $\Delta$ and the triplet bandwidth for $x<0.05$ \cite{Stone_2010}.  Therefore values of $J_0=16.42$ K, $J_1+J_4+2\left(J_2+J_3\right) = 5.31$ K and $g=2.07$ determined from the fit of the undiluted Ba$_3$Mn$_2$O$_8$ measurement (solid black circles) were held fixed for the fits of susceptibility data for the other samples.  These fixed values are in approximate agreement with the values measured from INS and EPR, of $J_0=18.78$ K, $J_1+J_4+2\left(J_2+J_3\right) = 11.03$ K and $g=1.96$, up to the inherent limitations of this fit.  The Curie constant C, extracted from these fits, is plotted against the left axis in Fig. \ref{Suscept}(b), and varies essentially linearly with V concentration.  The effective moment per mole of V, plotted against the right axis of Fig. \ref{Suscept}(b), is therefore independent of V concentration within the uncertainty of the measurement.  Observed values are close to those anticipated for Mn$^{5+}$, implying that each V impurity results in a single unpaired Mn$^{5+}$ spin.  That is, substitution of V does not appear to result in clustering, but rather, as anticipated in Fig. \ref{Struct}(c), the number of doubly-broken dimers is very small, and the majority of V impurities occupy half of a dimer site paired with a magnetic Mn$^{5+}$ ion.  The fit parameter $\alpha$ nominally scales with (1-$x$).  However, because the change in $\alpha$ is small relative to its absolute value for the low concentrations studied here, this provides a much less sensitive measure of the number of unbroken dimers than does the Curie susceptibility. The temperature independent background term $\chi_0$ was motivated by Langevin diamagnetism, but fit parameters were found to be weakly paramagnetic ($\chi_0 \cong 3 \pm 2 \times 10^{-4}$ emu/mol for the parent compound), most likely reflecting inaccuracies associated with the dimer model, and specifically the relatively crude mean-field approximation used to account for interdimer exchange interactions.  Uncertainty in fit values of $\chi_0$ were larger than any variation with composition.

\begin{figure}
\includegraphics[width=8cm]{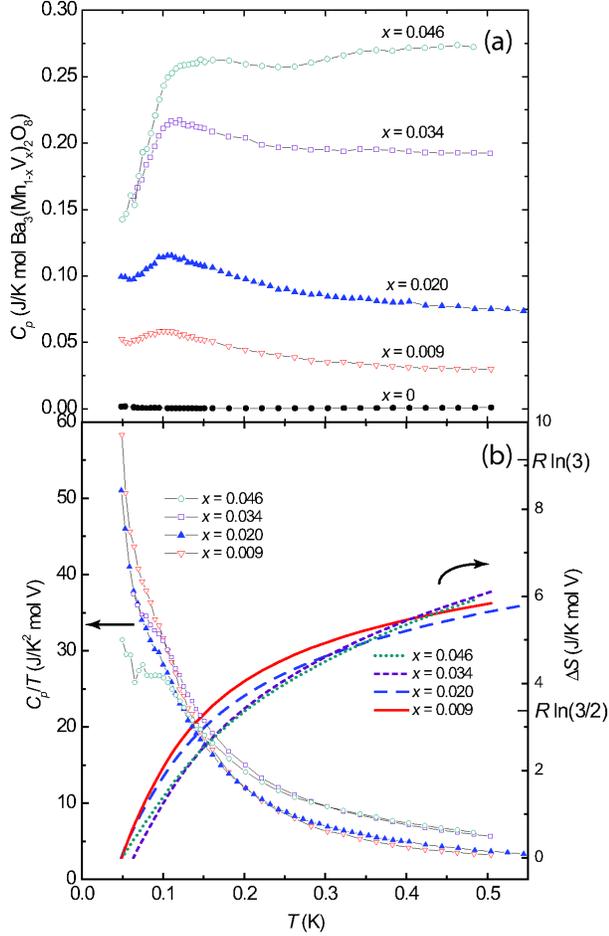}
\caption{(Color online) (a) Low temperature heat capacity of Ba$_3$(Mn$_{1-x}$V$_x$)$_2$O$_8$ in zero field.  Substitution of vanadium leads to a substantial increase in the total heat capacity.  (b) Left axis: The same data, plotted as $C_p/T$ and scaled by V concentration, revealing the evolution in the functional form of the heat capacity.  Right Axis: Change in entropy $\Delta S$ between 50 mK and 500 mK associated with the total heat capacity.  Despite subtle changes in the functional form of the heat capacity as the vanadium concentration is increased, the integrated entropy per mole of V over this temperature window remains remarkably similar.}
\label{Doping}
\end{figure}

\begin{figure}
\includegraphics[width=8.5cm]{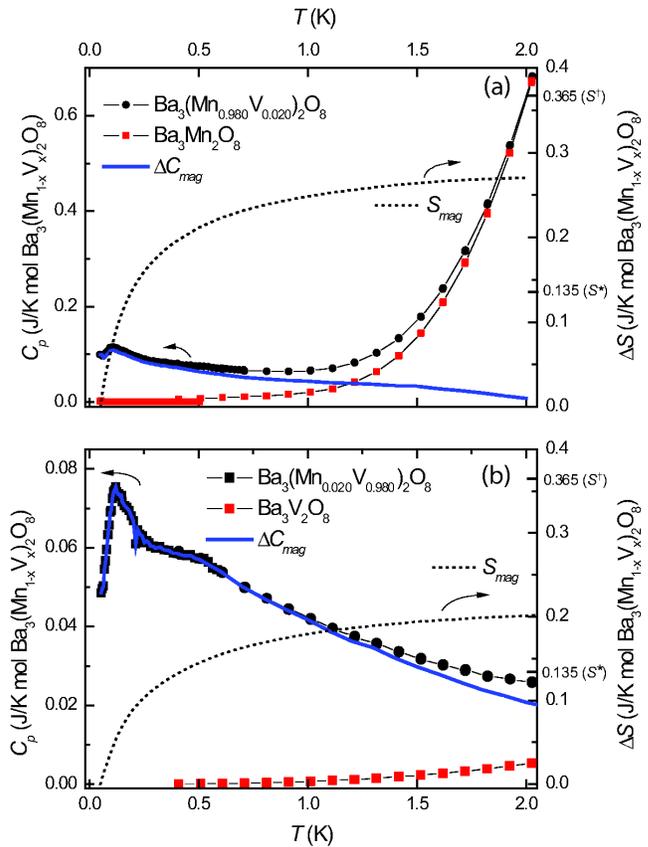}
\caption{(Color online) a) Heat capacity of Ba$_3$(Mn$_{0.980}$V$_{0.020}$)$_2$O$_8$, Ba$_3$Mn$_2$O$_8$ and the difference at 0 T shown in black circles, red squares and the blue line, respectively. The change in entropy calculated from the difference is plotted as a black dotted line along the right axis.  b) Heat capacity of Ba$_3$(Mn$_{0.020}$V$_{0.980}$)$_2$O$_8$, Ba$_3$V$_2$O$_8$ and the difference at 0 T shown in black circles, red squares and the blue line, respectively.  The change in entropy calculated from the difference is plotted as a black dotted line along the right axis.}
\label{Entropy}
\end{figure}

Heat capacity measurements were performed down to 50 mK using a dilution refrigerator.  Data are shown in Fig. \ref{Doping}(a) for the temperature interval 50 to 500 mK.   The stoichiometric parent compound, Ba$_3$Mn$_2$O$_8$ ($x=0$, black filled circles) exhibits only a very small heat capacity in this temperature interval due to the much larger energy scale of the spin gap.  A small temperature dependence due to the phonon contribution is barely discernable, but can be seen more clearly when similar data for Ba$_3$V$_2$O$_8$ are plotted over a wider temperature interval in Fig. \ref{Entropy}(b). In contrast, samples with $x>0$ reveal dramatically different behavior.  The heat capacity is approximately two orders of magnitude larger, and exhibits a broad maximum centered at approximately 110 mK. For $x$=0.009 and 0.020 there is a slight indication of an additional upturn in the heat capacity at the lowest temperatures.  This is not observed for the higher two V concentrations, $x$=0.034 and 0.046, which also exhibit a progressive broadening of the feature at 110 mK, and, for $x$=0.046, a gradual increase in the heat capacity from 200 mK to 500 mK.  Significantly, none of the concentrations studied exhibit a sharp anomaly characteristic of a phase transition.

The same data shown in panel (a) of Fig. \ref{Doping} are replotted as $C_p/T$ normalized by the vanadium concentration $2x$, in panel (b).  Scaled in this way, it is clear that $C_p$ approximately scales with $x$.  The change in entropy, $\Delta S$ over this temperature window can be estimated from the area under these curves, and is plotted against the right axis of panel (b), also normalized per mole of V.  The small phonon contribution has negligible effect on this estimate over this temperature range, as can be readily appreciated by inspection of the data in panel (a).  Inspection of panels (a) and (b) of Fig. \ref{Doping} reveals a progressive change in the T-dependence of the heat capacity and entropy as the vanadium concentration increases.  For the higher vanadium concentrations, relatively more entropy is removed per increment of temperature at high temperature than for lower concentrations, and vice versa at the lowest temperatures.  This concentration dependence is, however, rather subtle, and the most remarkable aspect of the data is that the total entropy removed over this interval is almost identical for all four compositions, despite a factor of 5 difference in the vanadium concentration.  This value [$\sim$ 6 J/K mole (V)] is approximately 60\% of the total magnetic entropy $S_{total}$ = $R \ln(3)$ (where $R$ is the molar gas constant) associated with the unpaired $S$=1 magnetic Mn ions induced by the vanadium substitution.

\begin{figure}
\includegraphics[width=8.5cm]{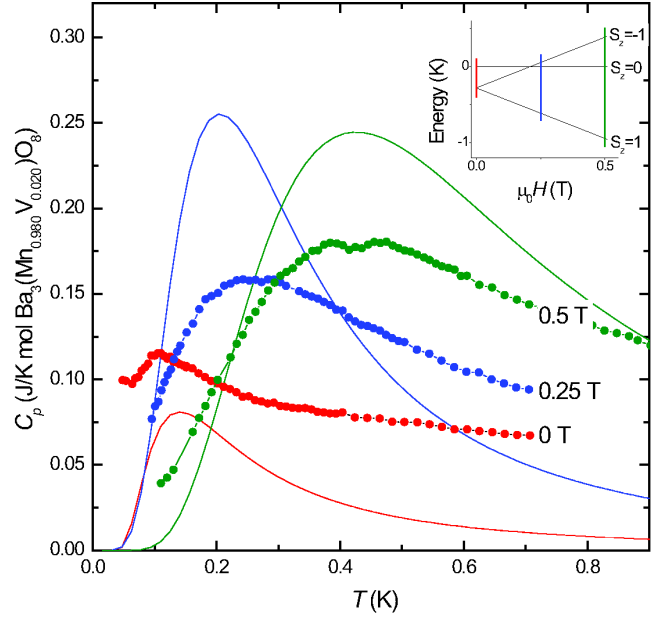}
\caption{(Color online) Heat capacity of Ba$_3$(Mn$_{0.980}$V$_{0.020}$)$_2$O$_8$ in magnetic fields of 0 T, 0.25 T and 0.5 T, shown as red, blue and green points, respectively.  Lines show the theoretical heat capacity for the Schottky anomaly associated with an isolated $S=1$ spin with single ion anisotropy $D=$ -0.024 meV in the same fields.  Inset shows the appropriate energy spectrum for this calculation.  Vertical lines indicate the three different fields for which measurements were performed.}
\label{Schottky}
\end{figure}

The heat capacity data for the V-substituted samples shown in Fig. \ref{Doping}(a) reveal a broad maximum centered at approximately 110 mK for all four compositions.  In the absence of any other interactions, uniaxial single ion anisotropy, represented by a term $D \left(S^z\right)^2$ in the spin Hamiltonian, will split the $S^z=0$ and $S^z=\pm1$ triplet states of unpaired Mn moments.  A calculation of the expected contribution to the specific heat in zero field is shown in Fig. \ref{Schottky} for the specific case of $x=0.020$ (red line), using the value of $D$ = -0.024 meV previously determined from EPR measurements of diluted Ba$_3$(Mn$_{1-x}$V$_{x}$)$_2$O$_8$.  The maximum value occurs at a temperature very close to that observed in experiment (solid red circles), indicating that the experimentally observed features are likely related to Schottky behavior.  To further test this hypothesis, the heat capacity was also measured in applied fields of 0.25 T and 0.5 T (green and blue data points in Fig. \ref{Schottky}) and compared with calculated values (blue and green lines).  The applied field splits the $S^z=\pm 1$ states (inset to Fig. \ref{Schottky}), resulting in broadening of the Schottky anomaly and a shift in the maximum value to higher temperatures.  The experimental data reveal very similar behavior, confirming that this feature is closely related to the Schottky behavior anticipated for unpaired Mn spins.

The correspondence between the calculated Schottky anomaly and the experimentally determined heat capacity is, however, not perfect.  Inspection of Fig. \ref{Schottky} reveals a significant discrepancy between the magnitude of the theoretical curves and the experimental data, even though the temperature at which the maxima occur agrees.  This difference is particularly striking for the 0 T data, for which the measured heat capacity is uniformly larger than the theoretical prediction, indicating that the magnetic entropy associated with the full triplet $S=1$ state is progressively removed over a fairly broad temperature interval.  For non-interacting moments, the doublet ground state should remain unsplit.  Hence the anticipated entropy associated with the Schottky anomaly in zero field is $S^{\star}=2x[R\ln(3)-R\ln(2)]=2xR\ln(3/2)$ (where 2$x$ is the number of moles of free Mn ions).  In contrast, the measured magnetic entropy between 50 mK and 500 mK (Fig. \ref{Doping}(b)) is considerably larger than this value.  Comparison of the measured heat capacity in zero field with the calculated Schottky behavior indicates that much of this difference occurs at a temperature considerably above the energy scale set by the single ion anisotropy.  Higher order terms in the crystal field expansion would only lead to splitting on a lower energy scale than the leading axial term and cannot account for this difference.  Nor does it seem likely that the progressive removal of entropy above 0.2 K is due to a spread of values of $D$ since the total entropy would still rise to $2xR\ln(3/2)$, and also since the anomaly centered at 110 mK is not especially broadened.  The data therefore indicate that interactions between the free moments play a significant role.  Indeed, the magnitude of the Schottky anomaly itself appears to be smaller than anticipated for the given concentration, but superimposed on top of a large and only weakly temperature dependent background, which must then arise from these interactions.  This is also borne out by measurements made in applied field.

In the presence of a magnetic field the $S^z=\pm1$ doublet is split, leading to an overall increase in the magnitude of the calculated heat capacity (blue and green lines in Fig. \ref{Schottky}).  For 0.25 T the measured heat capacity exceeds the calculated value for temperatures greater than 0.4 K, indicating the removal of magnetic entropy at higher temperatures, similar to the zero field data.  The area under the calculated blue and green curves is the full magnetic entropy of the free $S=1$ spins, corresponding to $S^{\dagger}=2xR\ln(3)$.  If some of this magnetic entropy is removed at higher temperature due to interactions between Mn moments, then there will be less magnetic entropy available at low temperatures, which is presumably why the measured heat capacity in 0.25 T falls below the calculated curve for temperatures below 0.4 K.  Data for 0.5 T appear to follow the same general form: more magnetic entropy is removed at higher temperature than would be anticipated for isolated Mn ions, reducing the magnitude of the low temperature Schottky anomaly.

Measurements of the heat capacity for $x=0.020$ were extended to higher temperatures in order to integrate the magnetic entropy over a wider temperature window (black data points in Fig. \ref{Entropy}(a)).  Two additional contributions to the entropy become significant above 0.5 K: the phonon contribution, and a magnetic contribution arising from thermal population across the spin gap associated with the intact dimer triplet states.  However, a reasonably accurate estimate of the magnetic heat capacity due to the unpaired Mn ions ($\Delta C_{mag}$, blue line) can be obtained by subtracting the heat capacity of the stoichiometric parent compound, Ba$_3$Mn$_2$O$_8$ (red squares).  The corresponding change in entropy $\Delta S$ between 50 mK and 2 K (right hand axis) is 0.291 J/K mol, substantially surpassing $S^{\star}=2xR\ln(3/2)=0.135$ J/K mol.

For comparison, the heat capacity was also measured over this temperature range for a crystal with $x=0.980$, containing approximately the same concentration of unpaired Mn moments as the $x=0.020$ sample.  For $x=0.020$ the small concentrations of V impurities introduces unpaired magnetic Mn ions, which interact via the singlet ``sea'' arising from the majority of unbroken dimers.  In contrast, for $x=0.980$, Ba$_3$V$_2$O$_8$ provides an ``empty'' magnetic background in which the Mn impurities interact.  Data are shown in Fig. \ref{Entropy}(b), together with an estimate of the magnetic contribution to the heat capacity ($\Delta C_{mag}$, blue line) obtained by subtracting the heat capacity of Ba$_3$V$_2$O$_8$ (red squares).  $\Delta C_{mag}$ rises with decreasing temperature, and appears to begin to curve over below approximately 0.5 K.  Below 0.2 K a clear Schottky anomaly is observed, with the maximum centered at 110 mK. The magnitude of the anomaly is comparable to that found for $x=0.020$, presumably for the same reason that a substantial amount of the total magnetic entropy is removed at higher temperatures.  The integrated entropy between 50 mK and 4 K is 0.271 J/K mol.  Measurements to higher temperature are hampered by systematic errors introduced by uncertainty in measurements of the sample mass for the two crystals used in these measurements.  However, it is clear that the integrated entropy over this window far exceeds that which would be estimated if the $S^z=\pm1$ doublet remains degenerate, $S^{\star}=0.135$ J/K mol.  Hence, for both $x$=0.020 and 0.980, it appears that interactions between unpaired Mn moments leads to progressive removal of magnetic entropy over a wide temperature range.  And in both cases a relatively small number of moments remain to a low enough temperature to contribute to the observed Schottky anomaly associated with the single ion anisotropy zero field splitting.  In neither case is there any evidence of a phase transition down to 50 mK.

\section{Discussion}

In the absence of long range magnetic order, it is reasonable to consider whether the low temperature state of Ba$_3$(Mn$_{1-x}$V$_x$)$_2$O$_8$ is a spin glass, the archetypal groundstate for systems with disorder.  In canonical spin glass systems, the temperature dependence of the heat capacity and magnetic entropy scales with moment concentration \cite{Wenger_1976, Martin_1980}.  Such systems exhibit a broad feature in the heat capacity, the maximum of which occurs at a temperature slightly above the freezing temperature, with roughly 80\% of the total magnetic entropy accounted for above $T_F$.  For example, in CuMn, an increase in the Mn concentration by an order of magnitude increases the freezing temperature by greater than a factor of 5, such that the magnetic entropy is collected over a much wider temperature range.  However, the behavior in Ba$_3$(Mn$_{1-x}$V$_x$)$_2$O$_8$ is quite different; in this case the entropy is collected over an equivalent temperature range independent of vanadium concentration for the range of compositions studied (Fig. \ref{Doping}(b)).  Thus it is unlikely that the groundstate is a canonical spin glass.  Equally, a Griffiths phase \cite{Griffiths_1969, Bray_1987} is not anticipated because the parent compound itself exhibits a singlet ground state, rather than long range magnetic order.

An alternative groundstate for disordered systems is the random singlet phase.  This state, which does not exhibit long range order, is characterized by formation of local singlets \cite{Ma_1979}.  Microscopically, as temperature is decreased the pair of most strongly coupled moments forms a singlet; then as temperature is further decreased the pair of remaining moments with strongest coupling forms a singlet, and so on until all the moments are paired.  This state was first proposed to account for the low-temperature susceptibility and heat capacity of lightly doped semiconductors, in particular Si:P \cite{Bhatt_1982, Bhatt_1992, Andres_1981, Kummer_1978, Lakner_1989}, and has since been of much theoretical and experimental interest for one dimensional disordered systems \cite{Fisher_1994, Fernandes_1994, Continentino_2001, Fernandes_2003, Monthus_1997, Saguia_2002}. The dominant thermodynamic signature of this phase is the susceptibility which diverges as $T^{\alpha}$ with $-1<\alpha<0$ at low temperatures (i.e a sub-Curie law), as observed in both Si:P, Cd:S and the one dimensional systems MgTiOBO$_4$ and MnMgB$_2$O$_5$ \cite{Andres_1981, Kummer_1978, Fernandes_1994, Fernandes_2003}.  Additionally, it was found that the heat capacity of Si:P increases with decreasing temperature for the lowest concentrations over an appreciable range of temperatures reflecting the progressive removal of magnetic entropy upon cooling \cite{Lakner_1989}.

A scenario in which the ground state of Ba$_3$(Mn$_{1-x}$V$_x$)$_2$O$_8$ is a random singlet phase is qualitatively consistent with the experimental observations.  Consideration of the relevant exchange energies serves to illustrate whether collective or local behavior is expected to dominate.   Specifically, a comparison of the maximum exchange energy between two impurities ($J_{max}\textbf{S}_i \cdot \textbf{S}_j$) with the average total exchange energy resulting from interaction with all of the neighboring impurities ($\sum_i Z_i P_{x,i} J_{ij} $, where $Z_i$ is the number of coordinating sites with exchange $J_{ij}$ and $P_{x,i}$ is the probability that these sites are occupied) distinguishes whether a random singlet phase or collective magnetic order (for example a spin glass) is favored.  For vanadium concentrations up to the maximum studied, $x=0.046$, it is straightforward to show that the random singlet phase is favored.  As an example, for the case of $x=0.046$ we first consider a free Mn moment for which the nearest unpaired Mn is a distance $r_2$ away (recall that $J_2$=0.256 meV is the strongest exchange interaction after $J_0$).  In comparison, the average sum of further interactions acting on the free Mn moment, which would be appropriate for a mean field treatment, is $3(0.046)J_1+6(0.046)J_3+3(0.046)J_4$ = 0.060 meV.  The clear energetic advantage of singlet formation continues when progressively smaller exchange couplings are considered after removal of all $J_2$-bonded pairs, and so on (a crude version of the ``decimation'' procedure first introduced to describe random singlet formation \cite{Ma_1979}).  This is due to the combined actions of the hierarchy of exchange interactions found in Ba$_3$Mn$_2$O$_8$, the small size of the Mn moment ($S=1$), and the low concentration of magnetic impurities.  This crude analysis clearly shows that the exchange between a single pair of spins dominates over more collective behavior, implying that a random singlet state is theoretically favored over a spin glass state, at least for the concentration regime considered here.

To further test this view, a model of the random singlet phase in Ba$_3$(Mn$_{1-x}$V$_x$)$_2$O$_8$ was undertaken in order to make a quantitative comparison with the measured heat capacity data.  The model is an approximation of the previously described ``decimation'' process \cite{Ma_1979}, with three changes: (1) we do not recalculate the probability of finding successively weaker random singlet pairs after each round of decimation (this leads to an overestimate of the probability of smaller pairing distances); (2) the exchange interaction between two unpaired moments forming a singlet is not renormalized by other exchanges between neighboring moments; and (3) we explicitly include the effect of single ion anisotropy.  Limitations of this model are discussed below.  The total heat capacity is calculated by summing over the probability that two moments with exchange $J(r)$ will pair, $P_x(J(r))$, times the heat capacity for two moments with exchange $J(r)$:
\begin{equation}
C_{p, tot} (T) = \sum_{J(r)} P_x (J(r)) C_p (J(r), T)
\end{equation}
Details of the heat capacity calculation for two moments with exchange $J$ are given in the Appendix.  The exchange between neighboring impurities $J(r)$ was determined by the distance between impurities. For neighboring impurities a distance $r\leq r_4$ apart, the effective exchange was approximated using the values determined from INS measurements of the undiluted system \cite{Stone_2008b}. For impurities at a distance $r>r_4$, the effective exchange was approximated by a decaying exponential as expected for localized moments: $J(r) = J'e^{-r/r'}$.  The parameters $J'$ and $r'$ were determined from fits of the known exchanges $J_1$ through $J_4$.  The probability $P_x(J(r))$ was determined by finding the distance between an impurity and the neighboring impurity with which it has the strongest exchange, and was calculated similarly to $P_x(r)$, shown in Fig. \ref{Struct}(c) \cite{ProbChange}.  For $r>r_4$, successive shells of widths $dr_n$ were taken, such that the number of atoms in each shell corresponds to a 5\% increase in probability (i.e. $P_x(J(r_n))$ = 0.05 where $r_n = r_{n-1}+dr_n$).

Both the estimated exchange model and also the probability functions used in this model are approximations. We have used ``bare'' exchange values, determined from INS measurements, but effective exchange values between unpaired Mn moments on broken dimers will be renormalized by the presence of the surrounding singlet states. In particular, geometric frustration naturally leads to a reduction in the effective exchange from the bare values \cite{EffEx}.  The model further approximates the exchange for $r>r_4$, in which the real effective exchange most likely does not vary perfectly exponentially for intermediate $r$ since the superexchange depends sensitively on bond angles and bond lengths.  Also, $P_x(J(r))$ overestimates the probability of smaller pairing distances.  This can be understood by considering an impurity $i$ with nearest neighboring impurity $j$.  If there is a third impurity $k$ which is closer to impurity $j$ than impurity $i$, then impurity $j$ pairs with impurity $k$ and not impurity $i$, implying that impurity $i$ pairs with a fourth impurity further away than impurity $j$, shifting probability to longer pairing distances.  Nevertheless, as shown below, this crude model appears to capture the observed behavior at a semi-quantitative level, justifying its consideration in terms of an initial description of the data.

\begin{figure}
\includegraphics[width=8.5cm]{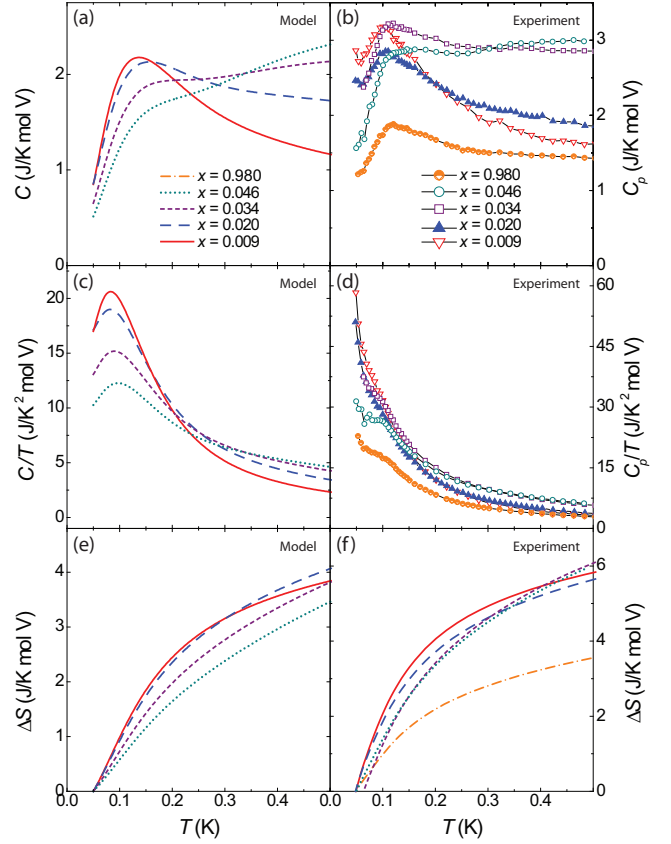}
\caption{(Color online) (Color online) Comparison of the experimentally observed heat capacity of Ba$_3$Mn($_{1-x}$V$_x$)$_2$O$_8$ with that of the random singlet model. Panels (a), (c) and (e) show $C$, $C/T$ and the integrated entropy $S$ respectively for the random singlet model, as described in the main text. Values have been scaled by the vanadium concentration. Panels (b), (d) and (e) show the experimental results for the same quantities. For $x$ = 0.980, the data have been normalized by the amount of Mn (i.e. 0.020) rather than the V concentration.}
\label{ModelCp}
\end{figure}

The calculated heat capacity, the heat capacity scaled by temperature and the entropy for the different vanadium concentrations experimentally studied are shown in Fig. \ref{ModelCp}(a), (c) and (e), respectively.  For $x\leq 0.046$, the calculated curves are in reasonable qualitative agreement with the experimental data, shown for comparison in panels (b), (d) and (f) (the experimental data have no background subtracted which is negligible in this temperature range as can be seen in Fig. \ref{Entropy}).  For $x=0.009$ and $x=0.020$, the model heat capacity increases as temperature decreases, reaching a maximum at the single ion Schottky peak before decreasing at lower temperatures.  For $x=0.034$ and $x=0.046$ the model heat capacity decreases monotonically as temperature decreases, with a shoulder at the single ion anisotropy Schottky peak.

The calculated entropy is also consistent with the experimental data (Fig. \ref{ModelCp}(e) and (f), respectively) for $x \leq 0.046$.  In spite of the five times difference in the vanadium concentration, which leads to a significant change in the probability distribution of exchange values of the random singlets, the change in entropy from 50 mK to 500 mK determined from the model rises to roughly the same value of $\sim$0.38 J/K mol V = 0.42 $R\ln(3)$ independent of vanadium concentration, in accordance with the experimental data.  By extending the model calculations to both lower and higher temperatures, the change in entropy for all vanadium concentrations eventually reaches the maximum available entropy, $R\ln(3)$.  However, the smaller vanadium concentrations recover most of the unaccounted entropy below 50 mK while the larger vanadium concentrations recover most of the unaccounted entropy above 500 mK.  In this sense it is somewhat coincidental that the experimentally accessible temperature window of 50 - 500 mK leads to an almost concentration independent change in entropy.

Despite the successful description of the evolution of the $T$-dependence of the heat capacity as a function of composition, both the calculated heat capacity and entropy are uniformly smaller than the experimental data, reflecting limitations of the simplified exchange model used for the calculations.  In particular, both the effective exchange $J(r)$ and the pairing probability $P_x(J(r))$ used in the calculations were approximations as described earlier.  As a consequence, the model heat capacity deviates slightly from the experimental data.  This can be seen most clearly by considering $C_p/T$, which emphasizes subtle differences at very low temperatures (Fig. \ref{ModelCp}(c) and (d)).  The random singlet model predicts a roll-over below $\sim$ 80 mK, whereas the experimental data continue to rise with decreasing temperature down to our base temperature of 50 mK.  This difference reflects the inaccuracy of the simplified model used to estimate the superexchange for the most distant pairs. However, the overall agreement is reasonable based on these approximations, lending considerable weight to the hypothesis that Ba$_3$(Mn$_{1-x}$V$_x$)$_2$O$_8$ realizes a random singlet phase at low temperature, at least for the range of concentrations studied in this report.

It is interesting to compare the heat capacity for $x$ = 0.020, corresponding to 2.0\% unpaired Mn ions in an otherwise ``filled'' Mn lattice, with that of $x$ = 0.980, corresponding to the case of 2.0\% Mn impurities in an otherwise ``empty'' lattice. Within the simplified random singlet model developed above, the heat capacity for the two cases is identical. However, as can be seen in Fig. \ref{ModelCp} (b), the measured heat capacity for these two compositions, although similar, is not identical. This difference reflects the renormalization of the bare superexchange values for the case of the filled lattice \cite{EffEx}, and indeed the data for $x$ = 0.980 conform more closely to the model (which uses the bare superexchange constants derived from INS measurements) than does the data for $x$ = 0.020. Nevertheless, both compositions qualitatively conform to the expectations of the random singlet model, exhibiting the removal of magnetic entropy over a broad range of temperatures and a modest feature associated with the single ion anisotropy.

\begin{figure}
\includegraphics[width=8.5cm]{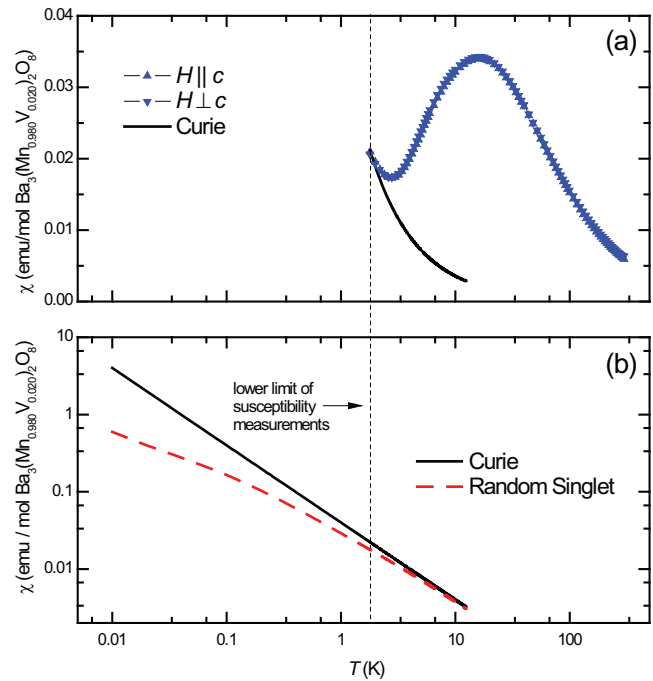}
\caption{(Color online) (a) Experimental susceptibility of Ba$_3$(Mn$_{0.980}$V$_{0.020}$)$_2$O$_8$ for fields both parallel (up triangles) and perpendicular (down triangles) to the $c$ axis.  Solid black line shows Curie contribution to the susceptibility determined from a fit of the entire data to Eq. \ref{SusEq}.  (b) Log-log plot comparison of susceptibility associated with concentration of unpaired Mn moments derived from the random singlet model (dashed red line) and Curie model (solid black line).  Vertical dashed line at 1.8 K marks the lowest temperature to which experimental measurements reached.  For the measured temperature range, the susceptibility of the random singlet model does not deviate substantially from Curie's law.}
\label{RSSuscept}
\end{figure}

Having introduced the random singlet model for the substituted lattice, it is worthwhile to briefly revisit the low-temperature susceptibility of Ba$_3$(Mn$_{1-x}$V$_x$)$_2$O$_8$. It is straightforward to show that the random singlet model results in a sub-Curie power law for the susceptibility, as anticipated \cite{Ma_1979, Bhatt_1982, Andres_1981, Kummer_1978, Fernandes_1994, Fernandes_2003} (see Appendix). The largest exchange interaction between unpaired Mn moments ($\sim J_2 \sim 0.256$ meV) sets the temperature scale below which singlet formation begins to occur. Hence, for temperatures above a few Kelvin, the calculated susceptibility tends towards a simple Curie behavior, as shown in Fig. \ref{RSSuscept}(b). For our experiments, susceptibility measurements were only possible down to 1.8 K, and thus in the analysis described earlier we used a simple Curie law to fit the data (Eq. \ref{SusEq}). As can be seen from Fig. \ref{RSSuscept}(a), the measured susceptibility is essentially isotropic over this temperature range, consistent with the small magnitude of the single ion anisotropy ($D \sim -0.024$ meV) and the nearly isotropic $g$-tensor. The observed upturn at low temperatures is not perfectly described by the Curie law, as anticipated from Fig. \ref{RSSuscept}(b), but experimental data extending to much lower temperatures are required for a quantitative comparison with the random singlet model.

Finally, we note that although the current data clearly indicate the absence of long range order down to 50 mK, we cannot rule out an ordering transition at a lower temperature.  However, the relatively small magnetic entropy that remains at this temperature, and the progressively larger physical separation of unpaired moments due to random singlet behavior, make long range order at a lower temperature rather unlikely.

\section{Conclusions}

In summary, the low-temperature properties of the site-diluted quantum magnet Ba$_3$(Mn$_{1-x}$V$_x$)$_2$O$_8$ have been explored through heat capacity experiments.  No sharp features associated with a transition into a ``order by disorder'' phase were observed, but rather we find a slowly varying magnetic contribution at low temperature due to the progressive removal of magnetic entropy over an extended temperature range. A Schottky anomaly induced by the single ion anisotropy is superimposed on this behavior below $\sim$250 mK.  There is very little variation in the temperature dependence of the magnetic entropy as vanadium concentration is varied, indicating that the behavior is not associated with spin freezing.  Rather, the data suggest that Ba$_3$(Mn$_{1-x}$V$_x$)$_2$O$_8$ has a random singlet ground state.

\section{Acknowledgements}

The authors thank C. D. Batista, D. Fisher, S. Haas and S. Kivelson for useful discussions.  We also acknowledge Robert E. Jones for technical assistance with electron microprobe measurements.  The work is supported by the DOE, Office of Basic Energy Sciences, under Contract No. DE-AC02-76SF00515.

\appendix

\section{Derivation of Heat Capacity and Susceptibility of a dimer including the effect of single ion anisotropy}

\begin{figure}
\includegraphics[width=8.5cm]{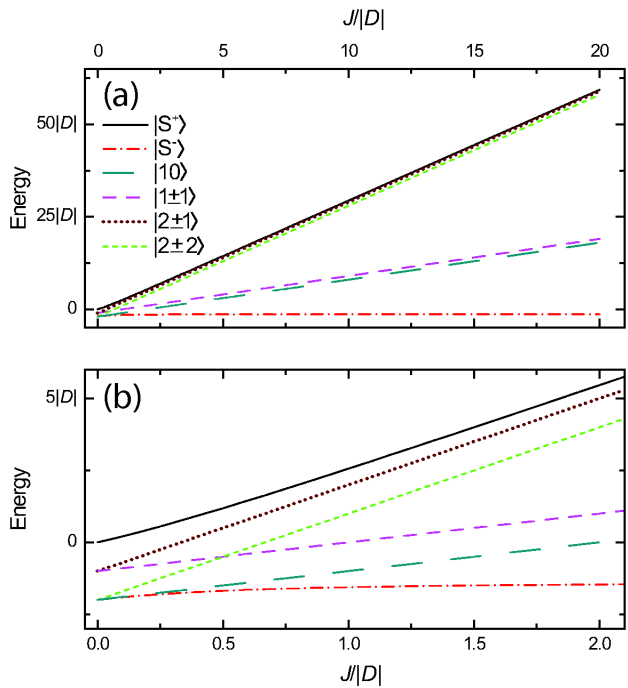}
\caption{(Color online) Energy spectrum of a pair of moments with interactions defined by Eq. \ref{RSEq} versus $J$, normalized by the single ion anisotropy $|D|$.  Large $J/|D|$ range and small $J/|D|$ range shown in (a) and (b), respectively.}
\label{Spectrum}
\end{figure}

Here we numerically calculate the heat capacity and susceptibility for a pair of spins with exchange $J$ and single ion anisotropy $D$.  This calculation is done for a range of antiferromagnetic exchange values relative to the easy axis single ion anisotropy, $J/|D|$.  The Hamiltonian acting on these two moments is:
\begin{equation}
\mathcal{H} =  J \textbf{S}_1 \cdot \textbf{S}_2 + D \left( (S_1^z)^2 + (S_2^z)^2 \right)
\label{RSEq}
\end{equation}
The energy spectrum for this system is most easily determined in the dimer basis.  In this case, all the dimer states are diagonal with the first term of the Hamiltonian, $J \textbf{S}_1 \cdot \textbf{S}_2$.  Further, the single ion anisotropy term $D \left( (S_1^z)^2 + (S_2^z)^2 \right) $ conserves total $S^z$ and cannot connect states of opposite symmetry, implying that only the $|00\rangle$ and $|20\rangle$ states can have off-diagonal matrix elements \cite{Samulon_2009}.  The matrix elements for all the states are:

\begin{eqnarray}
& \mathcal{H} | 20\rangle & =   \left( 3J  +  \frac{2D}{3} \right)|20\rangle + \frac{2\sqrt{2}D}{3}|00\rangle
\nonumber\\
& \mathcal{H} | 00\rangle & =  \frac{2\sqrt{2}D}{3}|20\rangle + \frac{4D}{3} |00\rangle
\nonumber\\
& \mathcal{H} | 2\pm2\rangle & = (3J + 2D)|2\pm2\rangle
\nonumber\\
& \mathcal{H} | 2\pm1\rangle & =  (3J + D)|2\pm1\rangle
\nonumber\\
& \mathcal{H} | 1\pm1\rangle & =  (J + D)|1\pm1\rangle
\nonumber\\
& \mathcal{H} | 10\rangle & =  (J + 2D)|10\rangle
\end{eqnarray}

Diagonalizing the $|00\rangle$ and $|22\rangle$ matrix elements results in eigenvectors $|S^{\pm}\rangle$ with eigenvalues $E_{|S^{\pm} \rangle}= \frac{1}{2} \left(2D+3J\pm \sqrt{\left(2D+3J\right)^2 -16DJ}\right)$.   The energy spectrum is plotted versus the size of the exchange $J/|D|$ in Fig. \ref{Spectrum}.  Panel (a) shows the spectrum for large values of $J/|D|$.  In this limit the system approaches the isolated dimer model, where the singlet has gaps of $J$ and $3J$ to the excited triplet and quintuplet states, respectively.  Panel (b) shows the spectrum for small values of $J/|D|$.  In this limit the system approaches the isolated moment model, where a dimer has total energy 0, $D$ or 2$D$ if it is composed of two, one or zero $S^z=0$ spins, respectively.

\begin{figure}
\includegraphics[width=8.5cm]{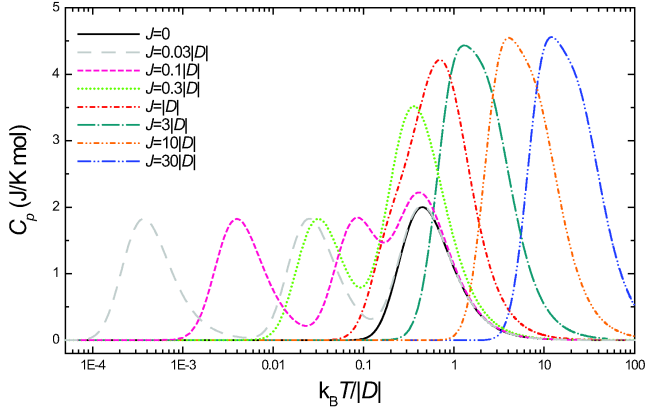}
\caption{(Color online) Heat capacity curves versus normalized temperature for specific $J/|D|$ values plotted on a log scale. Mol refers to one mole of spins.}
\label{DRange}
\end{figure}

The energy for a pair of moments with exchange $J$ can be expressed as a function of temperature:
\begin{equation}
E_J (T)  =  \frac{\sum_{i} E_{|\gamma_i\rangle} e^{-\beta E_{|\gamma_i\rangle} }}{\sum_{i} e^{-\beta E_{|\gamma_i\rangle} }}
\end{equation}
The summation runs over all the different basis states ($|\gamma_i \rangle = |S^\pm \rangle, |2 \pm2 \rangle, |2 \pm1 \rangle, |1 \pm1 \rangle, |1 0 \rangle $), $E_{|\gamma_i\rangle}$ is the energy of a given state and $\beta=1/k_B T$.  The heat capacity of a pair of moments with exchange $J$ is easily calculated numerically from the energy:
\begin{equation}
C_p (J, T) = \frac{d(E_J (T))}{dT}
\end{equation}
Several different heat capacity curves for a range of $J/|D|$ values are plotted on a log scale in Fig. \ref{DRange}.  For $J=0$ (solid black line), the system reduces to the isolated moment case and the heat capacity has a peak induced by the two gaps of the single ion anisotropy (a gap of $D$, $2D$ between the $|S^-\rangle, |10\rangle, |2\pm2\rangle $ ground states and the $|2\pm1\rangle, |1\pm1\rangle $ first excited states and $|S^+\rangle$ second excited state, respectively). For the largest $J$ values of $J=3|D|,10|D|$ and $30|D|$ there is a single peak centered at roughly 0.4$J$ arising from both the singlet-triplet and singlet-quintuplet gaps. For $J=|D|$ there is a single peak with a shoulder at lower temperatures.  For $J=0.3|D|$ there are split peaks, one centered roughly at the same position as the single ion anisotropy peak and a second at lower temperature.  Finally for the smallest non-zero $J$ values of $J=0.1|D|$ and $0.03|D|$ there are three peaks stemming from three gaps from the $|S^-\rangle$ groundstate: the gap to the first excited state, $|10\rangle$; the gap to the second excited states $|2\pm2\rangle$; and finally the single ion anisotropy gap to the third excited states, $|1\pm1\rangle,|2\pm1\rangle $ and $|S^+\rangle $.  The model shown in Fig. \ref{ModelCp} used a superposition of heat capacity curves taken from exchange values mostly within this range.  The different model curves used differently weighted superpositions, such that the higher (lower) vanadium concentrations are weighted more towards larger (smaller) $J$ values based on the probability distributions $P_x(J(r))$.

The susceptibility is calculated in the low field limit by similar means.  The calculation was done with $H \| c$ (i.e. along the dimer direction) for simplicity since most of the dimer states described above are diagonal and there is no appreciable anisotropy based on field direction for the temperature range studied (Fig. \ref{RSSuscept}(a)).  The low field susceptibility for a pair of moments with exchange $J$ as a function of temperature is:
\begin{equation}
\chi(J, T) = \frac{M(J, T)}{H} = \frac{1}{H}\frac{g \mu_B \sum_{i} m_{|\gamma_i\rangle}^z e^{-\beta E_{|\gamma_i\rangle} }}{\sum_{i} e^{-\beta E_{|\gamma_i\rangle} }}
\end{equation}
$m_{|\gamma_i\rangle}^z$ is the $z$ component of the angular momentum of energy eigenvector $|\gamma_i\rangle$, appropriate for fields along the $c$ axis.  The total susceptibility is calculated by summing over the probability that two moments with exchange $J(r)$ will pair, $P_x (r)$, times the susceptibility of two moments with exchange $J(r)$:
\begin{equation}
\chi_{tot}(T) = \sum_{J(r)} P_x(J(r))\chi(J(r), T)
\end{equation}

The susceptibility calculated from this random singlet model for an $x=0.020$ system (dashed red line) is plotted in a log-log scale in Fig. \ref{RSSuscept}(b) .


\begin{thebibliography}{1}
\bibitem{Prob1} The probability of finding at least one unpaired moment within a distance $r_i$ of a given unpaired moment, $P_{x, tot}(r_i)$, is determined by taking 1 less the probability that no unpaired moment is within a distance $r_{i}$ of a given a unpaired moment.   The probability that no unpaired moments are within a distance $r_i$ of a given unpaired moment is determined by multiplying the probabilities that there are no unpaired moments at each successive distance $r_j$ up to $r_i$ ($j \leq i$), taking account of the number of equivalent sites $Z_j$ in the $j$th shell: $1-P_{tot,x}(r_i) = \prod_{j=0}^i(1-x)^{Z_j}$
\bibitem{Prob2} The probability of the nearest unpaired moment being a distance $r_i$ away from a given unpaired moment, $P_x(r_i)$, is given by the probability that no unpaired moments are within a distance $r_{i-1}$ times the probability that there is an unpaired moment at a distance $r_i$, again taking account of the number of equivalent sites $Z_j$ in the $j$th shell: $P_x(r_i) = \left(\prod_{j=0}^{i-1}(1-x)^{Z_j} \right) \times \left(1-(1-x)^{Z_i}\right)$.
\bibitem{ProbChange} The relative hierarchy of exchange constants does not exactly follow that of the physical separation of sites, shown in Fig. \ref{Struct} (b). Specifically, $J_2>J_3>J_1>J_4$, whereas $r_1<r_2<r_3<r_4$. $P_x(J(r))$ is determined similarly to $P_x(r_i)$, the probability that the nearest unpaired moment is a distance $r_i$ away from a given unpaired moment: $P_x(J(r_i)) = \left(\prod_{j=0}^{i-1}(1-x)^{Z_j} \right) \times \left(1-(1-x)^{Z_i}\right)$.  The probabilities are calculated for a sequence of exchanges, with corresponding distances and coordination numbers, in which each successive exchange is smaller than all the previous exchanges.
\bibitem{EffEx} The origin of the reduction in the effective exchanges between impurities from the ``bare'' exchange values can be understood from an example. Two unpaired Mn moments connected by the $J_2$ exchange have an antiferromagnetic exchange with the third intact singlet dimer of the frustrated triangular structure. These antiferromagnetic exchanges induce an effective second order ferromagnetic exchange between the unpaired moments, reducing the overall value of the exchange between the unpaired moments.
\end{thebibliography}
\end{document}